\documentclass[usenatbib]{mnras}

\usepackage[T1]{fontenc}
\usepackage{aecompl}
\usepackage{amsmath,amsfonts,amssymb}
\usepackage[pdftex]{graphicx}
\usepackage{aas_macros}
\usepackage{color}
\usepackage{caption}

\title[Effects of photophoresis on the dust in a 3D PPD]{Effects of photophoresis on the dust distribution in a 3D protoplanetary disc}
\author[N. Cuello et al.]{N. Cuello$^{1}$, J.-F. Gonzalez$^{1}$ and F.C. Pignatale$^{1}$\\
$^{1}$Universit\'e de Lyon, F-69003 Lyon, France; Universit\'e Lyon 1, Observatoire de Lyon, 9 avenue Charles Andr\'e,\\ F-69230 Saint-Genis Laval, France; CNRS, UMR 5574, Centre de Recherche Astrophysique de Lyon;\\Ecole Normale Sup\'erieure de Lyon, F-69007 Lyon, France}
\begin{document}
\date{Accepted ... Received ...}

\pagerange{\pageref{firstpage}--\pageref{lastpage}} \pubyear{2015}

\maketitle

\label{firstpage}

\begin{abstract}
Photophoresis is a physical process based on momentum exchange between an illuminated dust particle and its gaseous environment. Its net effect in protoplanetary discs (PPD) is the outward transport of solid bodies from hot to cold regions. This process naturally leads to the formation of ring-shaped features where dust piles up. In this work, we study the dynamical effects of photophoresis in PPD by including the photophoretic force in the two-fluid (gas+dust) smoothed particle hydrodynamics (SPH) code developed by \cite{Fouchet2005}. We find that the conditions of pressure and temperature encountered in the inner regions of PPD result in important photophoretic forces, which dramatically affect the radial motion of solid bodies. Moreover, dust particles have different equilibrium locations in the disc depending on their size and their intrinsic density. The radial transport towards the outer parts of the disc is more efficient for silicates than for iron particles, which has important implications for meteoritic composition. Our results indicate that photophoresis must be taken into account in the inner regions of PPD to fully understand the dynamics and the evolution of the dust composition.

\end{abstract}



\section{Introduction}

	Thousands of extra-solar planets have been detected in the past 20 years \citep{Batalha2013}, which undoubtedly shows that planets are common byproducts of star formation. As a matter of fact, it is widely accepted that planets form in PPD around young stellar objects in a few Myr as the result of the accretion of kilometre-sized bodies called planetesimals \citep{ChiangYoudin2010}. These solids are formed through collisional growth of dust particles, with sizes ranging from $\sim$ 0.1 $\mu$m to $\sim$ 10~cm. Interferometric observations at different wavelengths probe these dust populations and suggest that grains grow in the disc \citep[and references therein]{Nattaetal2007}. Thus, in order to allow collisional growth to occur, dust grains must survive in the disc without being accreted by the central star.
	
	However, the study of the motion of solids inside the disc leads to the so-called radial-drift barrier \citep{W77}. A solid body orbiting a star in a Keplerian fashion loses angular momentum because of the aerodynamic drag of the gas, which orbits at a sub-Keplerian velocity. This is because the gas phase is pressure supported while the dust is not. The radial drift of a solid body is the most efficient when the Stokes number of a grain, which is the ratio between its stopping time and its orbital time, is close to 1. The stopping time is defined as the time it takes for a particle moving at a given initial velocity to reach the same velocity as the gas. For a particle of size $s$ and density $\rho_{\rm d}$ in the Epstein regime, the optimal size for radial drift in the midplane is given by \citep{Fouchet2010}:
\begin{equation} \label{eq:sopt}
 s_{\rm opt} (r) = \frac{\Sigma_{\rm g} (r)}{\sqrt{2 \pi} \rho_{\rm d}} \,\, ,
\end{equation}
where $\Sigma_{\rm g}$ is the gas surface density and $r$ is the distance to the star. $s_{\rm opt}$ depends on the considered disc model: for the Minimum Mass Solar Nebula (MMSN) models $s_{\rm opt}~\sim~1~{\rm m}$ at 1~au \citep{W77, H81}, while in classical T Tauri star protoplanetary discs $s_{\rm opt} \sim 1~{\rm dm}$ at 1~au \citep{Laibe2012}. If the radial drift is fast enough and the particle cannot decouple from the gas, then the solid body will be accreted by the star in time-scales of the order of hundreds of orbital periods. This is a severe problem for planet formation theory since it prevents dust grains to reach planetesimal sizes.

	Several mechanisms have been proposed to break the radial-drift barrier such as radial mixing \citep[and references therein]{Keller2004}; particle traps caused by planets \citep{Paardekooper2004, Fouchet2007, Fouchet2010, Ayliffe2012, Gonzalez2012, Pinilla2012, Gonzalez2015}; dead zones \citep{KretkeLin2007, Dzyurkevich2010, Armitage2011} and anticyclonic vortices \citep{Barge1995, Meheut2010, Regaly2012}; meridional circulation \citep{Fromang2011} and radiation pressure effects \citep{Vinkovic2014}. However, it remains unclear if a single mechanism or a combination of them can actually prevent accretion. In this work, we consider photophoresis as a possible mechanism to break the radial-drift barrier.
	
	Photophoresis has long been known \citep{Ehrenhaft1918}, however it is only recently that it has been considered from an astrophysical point of view \citep{KraussWurm2005}. This phenomenon is due to the temperature gradient across an illuminated dust particle in a low-pressure gaseous environment. In this configuration, a momentum exchange between the dust particle and the surrounding gas molecules takes place, which results in a motion of the particle away from the radiation source (cf. Section~2). In the optically thin regions of PPD dust particles are illuminated by the star and surrounded by low-pressure gas. Then, according to \cite{KraussWurm2005}, photophoresis might lead to the formation of ring-shaped dust distributions with sharp inner edges where particles pile up.
		
	However, it might be argued that photophoresis caused by stellar radiation only affects a small fraction of the disc. This is certainly true for young systems where the disc is still very massive and the stellar radiation cannot penetrate very deep into it. Nevertheless, in transitional discs at later evolutionary stages the amount of gas decreases as the gas dissipates and the inner regions become optically thin \citep{Augereau1999, Calvetetal2002, Wahhaj2005}.

	The fact that transitional discs tend to deplete the inner regions of gas in short time-scales could be seen as an issue. In fact, without gas the photophoretic effects completely vanish. However, \cite{WurmKrauss2006} suggest that there is a stage, just before the gas clearing, where the disc is still dense in gas but transparent for the stellar radiation. This phase might have existed in the late solar nebula. Considering transitional discs, \cite{TakeuchiKrauss2008} compute the location of the $\tau = 1$ line, which defines the transition between the optically thin and the optically thick regimes. By doing so, they estimate the locations in the disc which can be reached by stellar radiation. Given the small amount of $\mu$m-sized particles remaining at this epoch, they neglect the dust absorption and consider the Rayleigh scattering by H$_2$ as the main extinction mechanism. Their analysis leads to the important conclusion that these discs can be considered as optically thin for distances as large as 10 au (cf. their fig.~4). Considering one-dimensional transitional disc models, they computed, for different particle sizes, the equilibrium distances at which the photophoretic and the drag force cancel each other out. Their results indicate that photophoresis has a significant effect on the dust particles located from 0.01 au up to several au and that it might be able to stop their inward drift.
	
	Furthermore, microgravity experiments on illuminated mm-sized particles surrounded by gas show that photophoresis efficiently transports solids away from the radiation source \citep{Wurmetal2010}. Moreover, \cite{Duermann2013} measured the strength of the photophoretic force on illuminated plates with given porosities in low-pressure gaseous environments. Coupling their results to a one-dimensional disc model, they 
predicted that photophoresis would be able to stop the inward drift for large bodies with sizes ranging from a millimetre to a metre. Based on their measurements, they also demonstrated that porosity dramatically increases the photophoretic force felt by illuminated dust particles.
	
	The expression of the photophoretic force depends on the properties of the particle such as its size \citep{Rohatschek1995}, chemical composition \citep{LoescheWurm2012} and porosity \citep{Duermann2013}. Consequently, a radial sorting among grains of different properties may occur in the disc. This is experimentally demonstrated in \cite{Wurmetal2010} by measuring the photophoretic effects on different materials (glass, steel and chondrule-like particles). Their results show that photophoresis is less efficient for good heat conductors since the temperature gradient between the illuminated and the shadowed side of the particle is lower.
	
	The sorting of grains with different thermal properties may have played an important role in determining the observed properties of meteorites, such as carbonaceous chondrites, where high and low temperature material coexist in the same object \citep{ScottKrot2014}. Indeed, photophoresis naturally transports high-temperature forming material from the inner regions of PPD outwards \citep{WurmKrauss2006, Mousisetal2007}. The analysis of meteorites shows that, among the chondrites, most of the groups\footnote{A chondritic group contains chondrites with similar chemical properties, presumably formed from the same parent body.} are depleted in iron relative to bulk solar system abundances (CI chondrites). However, there are some groups, like EH and EL, which show an overabundance of iron \citep{ScottKrot2014}. This indicates that a mechanism has to operate in the inner regions of the solar nebula in order to separate iron-rich and silicate particles. \cite{Moudensetal2011} computed the efficiency of the photophoretic transport for different grain sizes over the disc lifespan in one-dimensional MMSN models including turbulence. They found that photophoresis leads to accumulation regions in the disc in agreement with the ring-shaped dust distributions first proposed by \cite{KraussWurm2005}. \cite{McNally2015} computed the photophoretic force on opaque spherical particles in a dilute gas in the optically thick regime. In this case, the temperature gradient is caused by the energy dissipation in the accretion disc, which is highly localized and results in strong vertical temperature fluctuations \citep{McNally2014}. Then, non-illuminated dust particles can be locally transported by photophoresis in the vertical direction. However, since in this work we focus on optically thin transitional discs, this effect will not be considered here.
	
	To date, no studies have explored the complex effects of photophoresis in PPD through hydrodynamical simulations. The aim of this paper is twofold: on the one hand, we study the efficiency of the outward transport of solids due to photophoresis and, on the other hand, we attempt to connect the resulting dust dynamics with the chemical composition of the grains. We give an overview of the photophoretic force in Section~2, detail the method in Section~3, show our results in Section~4, discuss the radial transport due to photophoresis in Section~5 and conclude in Section~6.

\section{The photophoretic force}
	
	\subsection{A model for all pressures}

\begin{figure}
\centering
\includegraphics[width=60mm]{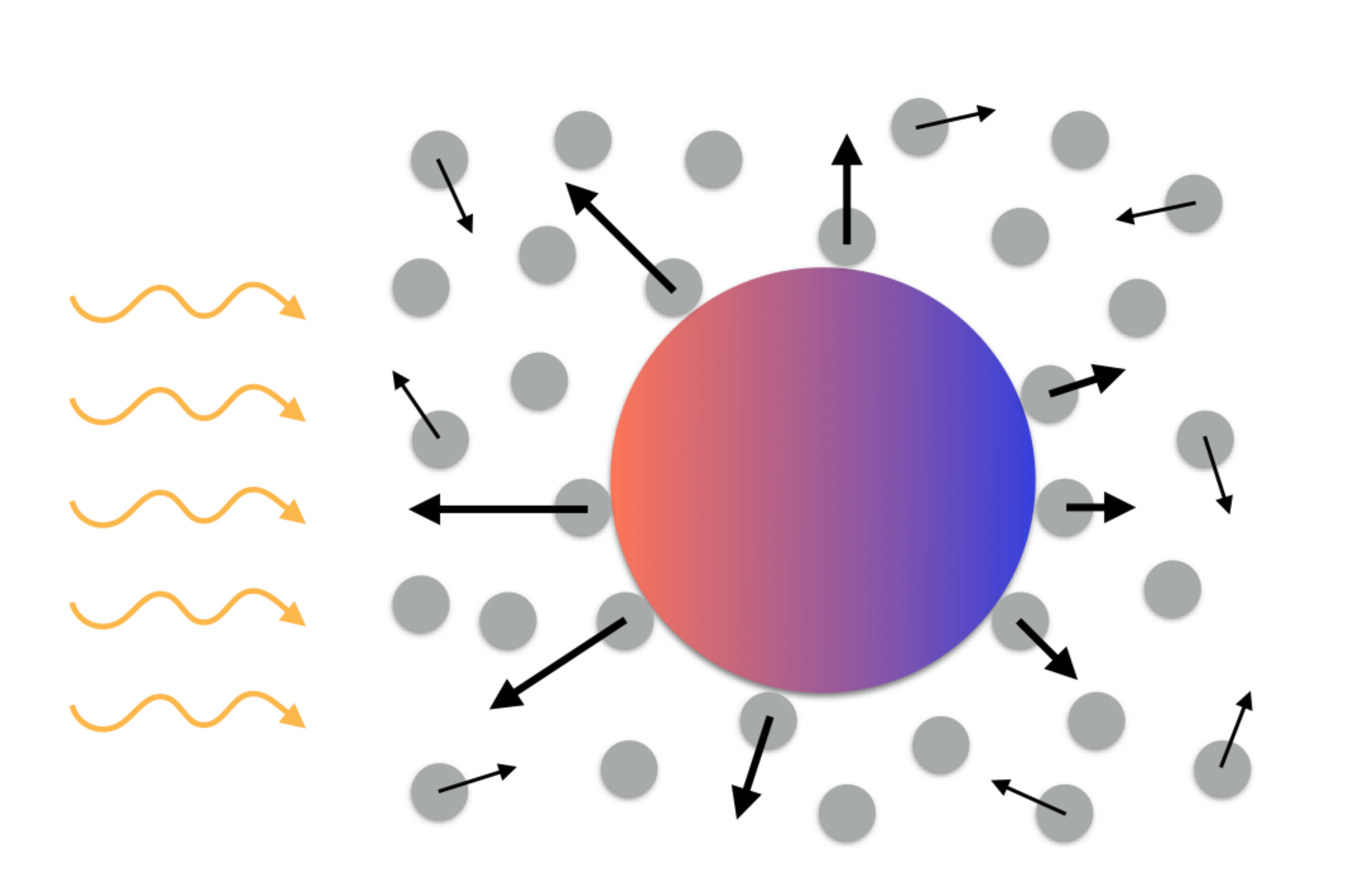}
 \caption{Momentum exchange between the dust grain and the gas particles (radiation from the left).}
 \label{fig:momenta}
\end{figure}

\begin{figure}
\centering
\includegraphics[width=60mm]{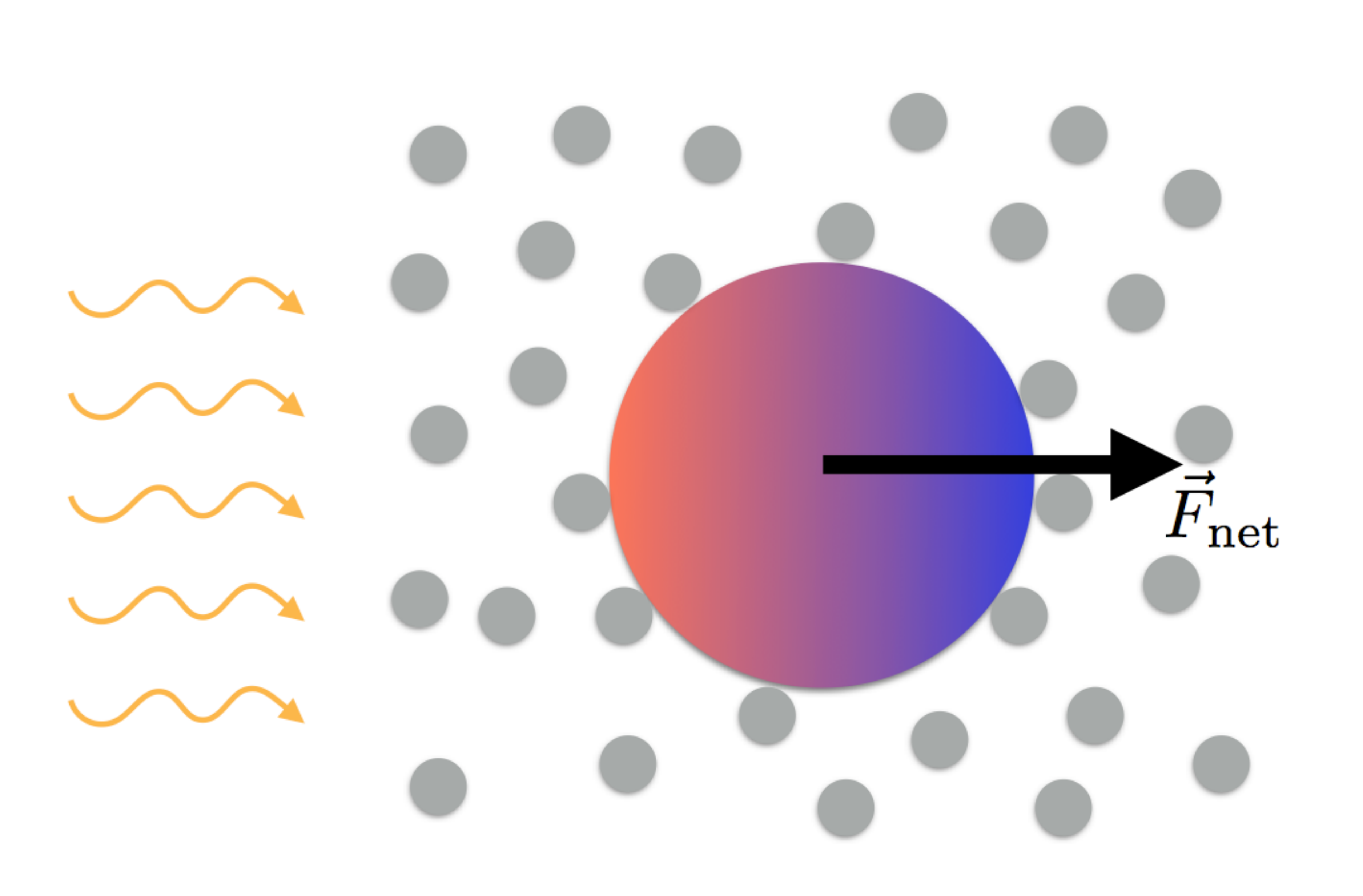}
\caption{Net force on the dust grain (radiation from the left).}
\label{fig:netforce}
\end{figure}

	The sketch in Fig.~\ref{fig:momenta} shows an illuminated dust particle embedded in a low-pressure gaseous environment with the radiation coming from the left. In this configuration, a temperature gradient develops on its surface: the hotter side is on the left and the cooler side is on the right. We assume that all the gas molecules have the same mass and thermal velocity close to the dust grain. When these molecules are diffusively scattered\footnote{Diffusive scattering does not include specular reflection.} by the particle surface, they equilibrate and reach the same temperature. Thus, their new velocities depend on where they hit the surface: molecules leaving the hotter side will have larger velocities compared to the ones on the cooler side. Therefore, the momentum exchange between the particle and the gas environment is anisotropic and results in a net force on the dust grain directed away from the radiation source as shown in Fig.~\ref{fig:netforce}.
	
	The rotation of the dust particles could lower the temperature gradient over the surface of the grain and consequently the intensity of the photophoretic force. \cite{LoescheWurm2012} identified three main mechanisms that can lead to rotation of the particles in the disc: random rotation processes, collisions with other grains and Brownian motion. However, under the typical conditions of pressure and density in PPD, the time-scales for rotation are several orders of magnitude longer than the time-scale for thermal conduction \citep{LoescheWurm2012}. Hence, the temperature gradient over the grain surface re-establishes itself much faster than it can be suppressed by rotation. This indicates that we can safely neglect rotation effects on the photophoretic force.
	
	
	The expression for the photophoretic force, $F_{\rm ph}$, used in this work is derived in \cite{Rohatschek1995} through the study of illuminated spherical particles in aerosols. It is a semi-empirical formula which covers the entire range of pressures, from the free molecular flow regime at low pressure to hydrodynamic flow at high pressure. There are three regimes for the photophoretic force according to the pressure, $p$, of the gaseous environment: the low-pressure regime for which $F_{\rm ph} \propto p$, the high-pressure regime for which $F_{\rm ph} \propto 1/p$, and the transition regime when $F_{\rm ph}$ reaches its maximum. These regimes are defined by the value of the Knudsen number defined as ${\rm Kn} = \lambda/s$, where $\lambda$ is the mean free path of the gas molecules and $s$ is the radius of the solid particle.  Its expression is given by $\lambda = k_{\rm B} T / \sqrt{2} \pi d^2 p$, where $k_{\rm B}$ is the Boltzmann constant and $d$ is the diametre of the gas molecules. In the low-pressure regime $\rm{Kn} \gg 1$, whereas in the high-pressure regime $\rm{Kn} \ll 1$. The expression for the photophoretic force reads as follows:
\begin{equation} \label{eq:Fph}	
 F_{\rm ph} = \frac{ 2 F_{\rm max}}{p / p_{\rm max} + p_{\rm max} / p}
 \end{equation}
with
\begin{equation}
	F_{\rm max} = \frac{s^2}{2} D \sqrt{\frac{\alpha}{2}} \frac{I}{k} \label{eq:Fmax} \,\,\, ,
\end{equation}
\begin{equation}
 	p_{\rm max} = \frac{3T}{\pi s} D \sqrt{\frac{2}{\alpha}} \label{eq:pmax} \,\,\, , 
\end{equation}
\begin{equation}
 	D = \frac{\pi \bar c \eta}{2T} \sqrt{\frac{\pi \kappa_{\rm t}}{3}} \label{eq:D} \,\,\, ,
\end{equation}
and
\begin{equation}
	\bar c = \sqrt{\frac{8 R_{\rm g} T}{\pi \mu}} \label{eq:c}
\end{equation}
where $I$ is the irradiance of the incident beam of light, $\alpha$ is the thermal accommodation coefficient (dimensionless and taken equal to 1 \citep{Rohatschek1995}), $k$ is the thermal conductivity of the solid particle, $T$ is the gas temperature, $\eta$ is the viscosity of the gas, $\kappa_{\rm t}$ is the thermal creep coefficient (equal to 1.14 \citep{Rohatschek1995}), $R_{\rm g}$ is the universal gas constant and $\mu$ is the molar mass of the gas. All the quantities on which Eqs. \eqref{eq:Fmax} to \eqref{eq:c} depend are functions of the disc local conditions and the distance to the star.

	The photophoretic force is often split into two components: the $\Delta T$-force mainly driven by the temperature gradient between the illuminated and shadowed sides of the solid particle, and the $\Delta \alpha$-force which depends on the inhomogeneities in composition and shape of the particle. In this work we consider spherical and homogeneous dust particles, so we neglect the $\Delta \alpha$-term and  take constant values for $\alpha$ and $\kappa_{\rm t}$. Since the dust in the disc is composed of different species, it is particularly interesting to study how the  \mbox{$\Delta T$-term} varies with the chemical composition and the size of the solid bodies.

	\subsection{Effects of porosity on thermal conductivity}

	As already mentioned, porosity is an important parameter given that grain growth strongly depends on the compositions and relative velocities of the impacting bodies \citep{Blum2010}. Laboratory experiments support the fact that collisions between submm- and mm-sized ${\rm SiO_2}$ aggregates result in dust grains with high porosities \citep{Meisneretal2013}. In addition, \cite{Kataoka2013} computed the evolution of porous dust grains in MMSN-like disc models and found that porosity has strong effects on their radial motion. In fact, porous grains grow more efficiently compared to compact grains, which eventually allow the former to decouple from the gas and prevent accretion.
	
	Porosity also affects the thermal properties of the particles. At a microscopic level, heat conduction is treated as the propagation of packets of vibrational energy, called phonons, through a crystal lattice \citep{Presley1997}. For solids bodies ranging from millimetre to decimetre sizes made of aggregated grains, the heat conduction strongly depends on the contact surface between these grains. Since a porous body has a smaller contact surface compared to a bulk solid, the resulting thermal conductivity is significantly lower \citep{Friedrichetal2008}. Therefore, an increase in porosity translates into an increase in the magnitude of the photophoretic force given by Eq.~\eqref{eq:Fmax}.

	Due to the lack of thermal conductivity data for large cm- and dm-bodies, it is difficult to fix a value for $k$ which would simultaneously depend on size, composition, porosity and temperature. In this context, porosity can be defined as the volume of empty space in a given particle divided by the total volume of the particle. The thermal conductivity for bulk silicates is of the order of magnitude of 1 $\rm{W\,m^{-1}\,K^{-1}}$ \citep{Opeil2012}, while if we consider the same silicates with a few percent of porosity, $k$ drops by at least one order of magnitude \citep{LoescheWurm2012}.
		
\begin{figure}
\includegraphics[width=90mm]{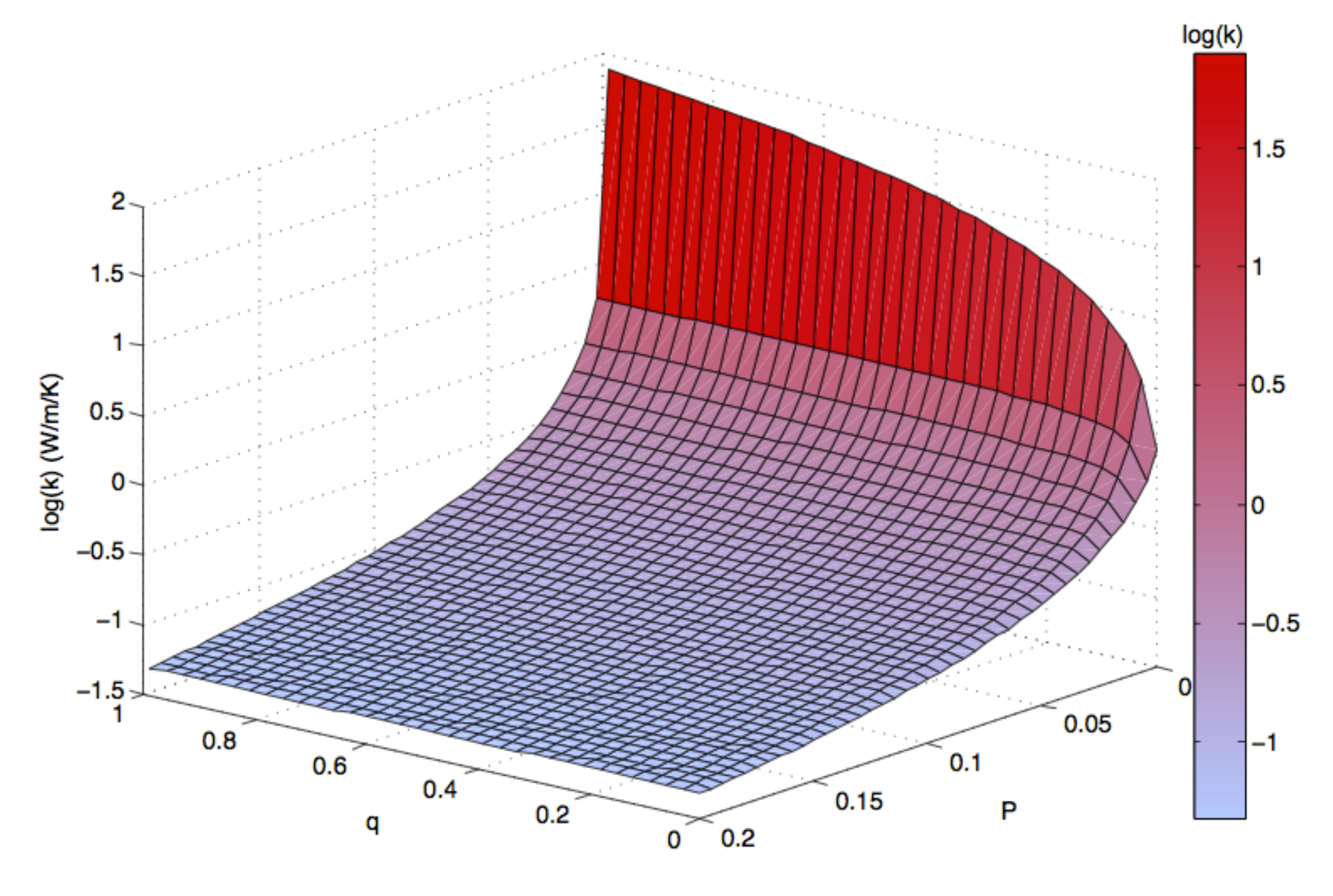}
\caption{Thermal conductivity for a porous dust particle at $T=280$ K with our 1D model. The colour bar corresponds to the logarithm of the thermal conductivity.}
\label{fig:thermalcond}
\end{figure}

	A very simple one-dimensional model allows to understand the effect of porosity on thermal conductivity. As a first approximation, we consider a porous solid body of length $L_0$ as a superposition of $n$ layers of length $l_j$ and thermal conductivities $k_j$. This is equivalent to a composite solid body of length $L_0$ with a total thermal resistance, $R$, equal to the sum of the thermal resistance of each layer. The thermal conductivity, $k$, is defined as the length divided by the thermal resistance: $k=L_0/R$. Thus, the resulting thermal conductivity for the porous particle is given by:
\begin{equation}\label{eq:effective_k}
	k = \frac{L_0}{\sum_{j=1}^{n} l_j / k_j} \;\;\;.
\end{equation}
We can then represent the pores inside the dust particle as voids filled with gas at the surrounding temperature. In this case, we obtain alternate layers of dust and gas with thermal conductivities $k_{\rm d}$ and $k_{\rm g}$ respectively. If we define the porosity $P$ as the length corresponding to pores divided by the total length of the particle, we can express Eq. (\ref{eq:effective_k}) in the following form:
\begin{equation}\label{eq:effective_k2}
	k = \frac{L_0}{ \frac{P L_0}{k_{\rm g}} + \frac{(1-P) L_0}{k_{\rm d}} } = \frac{k_{\rm d} k_{\rm g}}{P(k_{\rm d}-k_{\rm g}) + k_{\rm g}} \;\;\;.
\end{equation}
As an illustrative example, we consider a solid body made of a mixture of silicate and iron at 280 K, i.e. the fiducial temperature at 1 au in the MMSN models. We define the iron to silicate ratio, $q$, as the ratio between the amount of iron and the amount of silicate by mass. In this case, $k_{\rm d}= q \, k_{\rm iron} + (1-q) \, k_{\rm sil}$, where $k_{\rm iron}$ and $k_{\rm sil}$ are the thermal conductivities for iron and silicate particles respectively. According to tables in \cite{Touloukian1970}, at this temperature: $k_{\rm g} = 0.01\,\, \rm{W\,m^{-1}\,K^{-1}}$, $k_{\rm sil}= 1.147\,\, \rm{W\,m^{-1}\,K^{-1}}$ and $k_{\rm iron}= 80\,\, \rm{W\,m^{-1}\,K^{-1}}$. We let $q$ vary from $0$ to $1$, i.e. from pure iron to pure silicate particles. The resulting thermal conductivity is plotted in Fig.~\ref{fig:thermalcond} as a function of $P$ and $q$. A small amount of porosity is enough to dramatically decrease the thermal conductivity: for instance, if $P \geq 10$ \% then $k \leq 0.1$ $\rm{W\,m^{-1}\,K^{-1}}$, regardless the composition of the particle. Despite this simplified approach, these values are in agreement with the thermal conductivities measured for small porous chondrules \citep{Opeil2012}. However, our model might not be appropriate in a very low pressure environment where there is no gas within the pores. In this case, the heat transfer in pores might be dominated by radiation or, if the pores are small, by conduction through the dust itself \citep {Presley2010a, Presley2010b}. Other studies have been recently conducted considering two-layered particles made of a compact core covered by a fine-grained rim with a lower conductivity: \cite{LoescheWurm2012} and \cite{McNally2015} solve the heat equation in order to compute the resulting conductivity and find that the outer porous layer dramatically decreases the thermal conductivity of the bulk particle. These detailed calculations are also in agreement with our estimate for the thermal conductivity. This fact motivates our choice of a constant thermal conductivity for iron and silicate particles. We choose to take $k=0.1\,\, \rm{W\,m^{-1}\,K^{-1}}$, which is equivalent to considering particles with an average porosity of the order of 10\%.
	
\section{Numerical simulations}

	\subsection{Code description}
	
	We study the effect of photophoresis in PPD by means of the Lagrangian code of \cite{Fouchet2005}. We treat the gas and  the dust as two separate fluids interacting through aerodynamical drag. The code solves the equations of motion for each phase through the SPH formalism. Our goal is to assess the efficiency of the radial transport due to photophoresis. We add the photophoretic force to the equation of motion for the dust particles, which contains in addition the stellar gravity and the gas drag term. The equation of motion for the gas particles also contains three terms: the gas pressure, the stellar gravity and the back-reaction drag due to the interaction with the dust phase. \cite{Fouchet2005} gives a detailed description of the code and its limitations.  The criterion for the Epstein regime, $\lambda \geq 4 s / 9$, establishes the upper limit for which we can compute the gas drag using the equation:
\begin{equation}\label{eq:drag}
	F_{\rm d} = \frac{4 \pi}{3} \rho_{\rm g} c_{\rm s} s^2 \Delta v \,\,\, ,
\end{equation}
where $\rho_{\rm g}$ is the gas density, $c_{\rm s}$ the sound speed and $\Delta v$ is the differential velocity between dust and the mean gas motion. For the type of disc considered in this study this limit is approximately equal to 1 metre (fig. 5 in \cite{Laibe2012}), therefore the drag for particles with $ s \leq 1$ m can be computed using Eq.~\eqref{eq:drag}. For larger sizes, we enter the Stokes regime and this equation is not valid anymore.

	As explained in Section 2, the photophoretic force for a given particle of index $i$, $F_{\rm ph,i}$, depends upon the dust grain properties, the local properties of the disc and the distance to the star. These quantities are known for any given particle of our simulations. Thus, $F_{\rm ph,i}$ can be straightforwardly calculated by using Eq. (\ref{eq:Fph}) as follows:
\begin{equation} \label{eq:FphSPH}	
	F_{{\rm ph},i}(r_i,s_i,T_i, \left< \rho_{{\rm g},i} \right>) = \frac{ 2 F_{{\rm max},i}}{\left< p_i \right> / p_{{\rm max},i} + p_{{\rm max},i} / \left< p_i \right>} \;\;\;,
 \end{equation}
 where $\rho_{{\rm g},i}$ is the gas density and $r_i$ is the distance to the star for particle $i$. $F_{\rm max}$ depends on the radiant flux density $I(r_i)$ which is simply computed as the luminosity divided by the surface of the sphere of radius $r_i$, and the quantities in brackets are computed through SPH interpolations. 
 
 	We apply the SPH prescription for the viscosity in the disc with $\alpha_{\rm SPH} = 0.1$ and $\beta_{\rm SPH} = 0.2$, which control the strength of the viscosity \citep{Fouchet2007}. \cite{Arena2013} demonstrated that the SPH formalism adequately reproduces the turbulence expected in the disc. The chosen values for $\alpha_{\rm SPH}$ and  $\beta_{\rm SPH}$ correspond to a uniform value of the \cite{ShakuraSunyaev} parameter $\alpha \sim 0.01$, as suggested by observations of PPD \citep{King2007}.
 
 	It is worth noting that, instead of taking a constant value for the gas viscosity $\eta$, we compute it in the following way:
\begin{equation}\label{eq:viscosity}
	\eta(r_i) = \frac{\left<\rho_{\rm g}\right> \, \lambda \, c_{\rm s}(r_i)}{2} \;\;\;,
\end{equation}
where $c_s$ only depends on $r_i$ in the case of a vertically isothermal disc. Since the gas pressure at $r_i$ also depends on the height above the midplane, the photophoretic force varies with height in our simulations.

	\subsection{Disc model}

	We choose to study the same transitional disc model as \cite{TakeuchiKrauss2008} to be able to consider the inner regions of the disc as optically thin (cf. Section~1). The disc extends from 0.01 to 100 au and has a total mass of gas equal to $M_{\rm g} = 1.2 \times 10^{-2} \, M_\odot$. It orbits a $1 \, M_\odot$ star of $1 \, L_\odot$ luminosity and it is composed of 99\% gas and of 1\% dust by mass. The disc is non-self-gravitating, geometrically thin and vertically isothermal where the temperature follows the radial power-law $T(r) = 280 \, (r / 1\,{\rm au})^{-1/2}$ K. The surface density for the gas also follows a power-law $\Sigma_g(r) = 25 \, (r / 1\,{\rm au})^{-1/2}$ g cm$^{-2}$. The vertical density profile is a Gaussian profile $\rho(r,z) = \rho(r,0) \exp(-z^2/2 h^2)$, where $h$ is the disc scale height. This density profile corresponds to the vertical hydrostatic solution for a thin disc. We set $h(r=1\, {\rm au}) = 3.33 \times 10^{-2}\, {\rm au}$ and we have $h(r) \propto r^{5/4}$. This density profile is taken as the initial state of the disc for our simulations. We only focus in the inner parts of the disc from 0.1 to 5 au since this is where we expect photophoretic transport to be more efficient \citep{TakeuchiKrauss2008}.
	
\begin{figure}
\includegraphics[width=8cm]{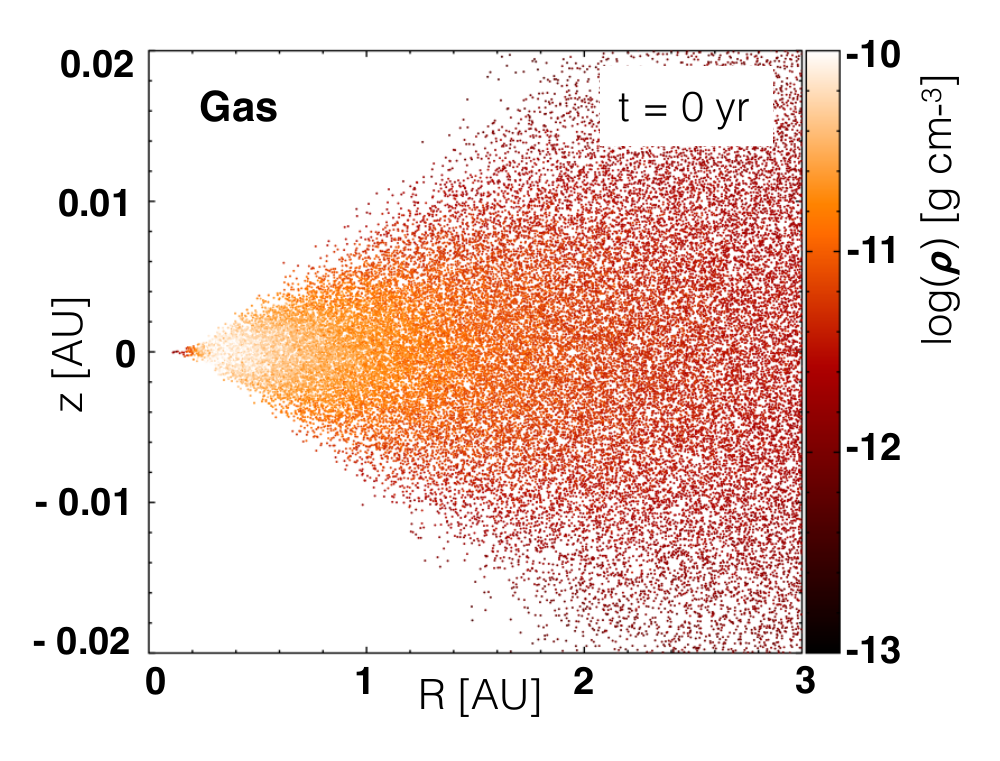}
\caption{Meridian plane cut of the gas distribution at the end of the relaxation phase. The coulour bar represents the logarithm of the volumetric density of the fluid.}
\label{fig:gasdisc}
\end{figure}
	
	\subsection{Simulation setup}
		
	We start with 100.000 gas particles distributed in such a way that we obtain the expected profile for the density (cf. Section 3.2). The initial velocity of the gas particles is set equal to the Keplerian velocity. Particles moving interior to 0.1 au and outside 5 au are considered as lost. After a relaxation phase of approximately 16 yr, the gas disc reaches an equilibrium state (Fig.~\ref{fig:gasdisc}). We then inject an equal number of dust particles on top of the gas particles with the same velocity. This state constitutes the $t=0$ configuration for all the simulations. We let the system evolve for approximately 250 years, i.e. 250 orbits at 1~au. We run a series of 10 simulations with 200.000 SPH particles to study the dependence of photophoresis on the intrinsic density (silicates, iron) and on the size of the solid bodies (1~mm to 1~m) as shown in Table \ref{tab:simus}. P and NP stand for simulations with and without photophoresis respectively. By varying the intrinsic density we are able to simulate different dust species: we take $\rho_d^{\rm sil} = 3.2$ g~cm$^{-3}$ and $\rho_d^{\rm iron} = 7.8$ g~cm$^{-3}$ for the silicate and the iron particles respectively. Given the Lagrangian nature of the SPH formalism, we are able to track the motion of the SPH particles in 3D throughout the disc evolution.
	
\begin{figure*}
\includegraphics[width=16cm]{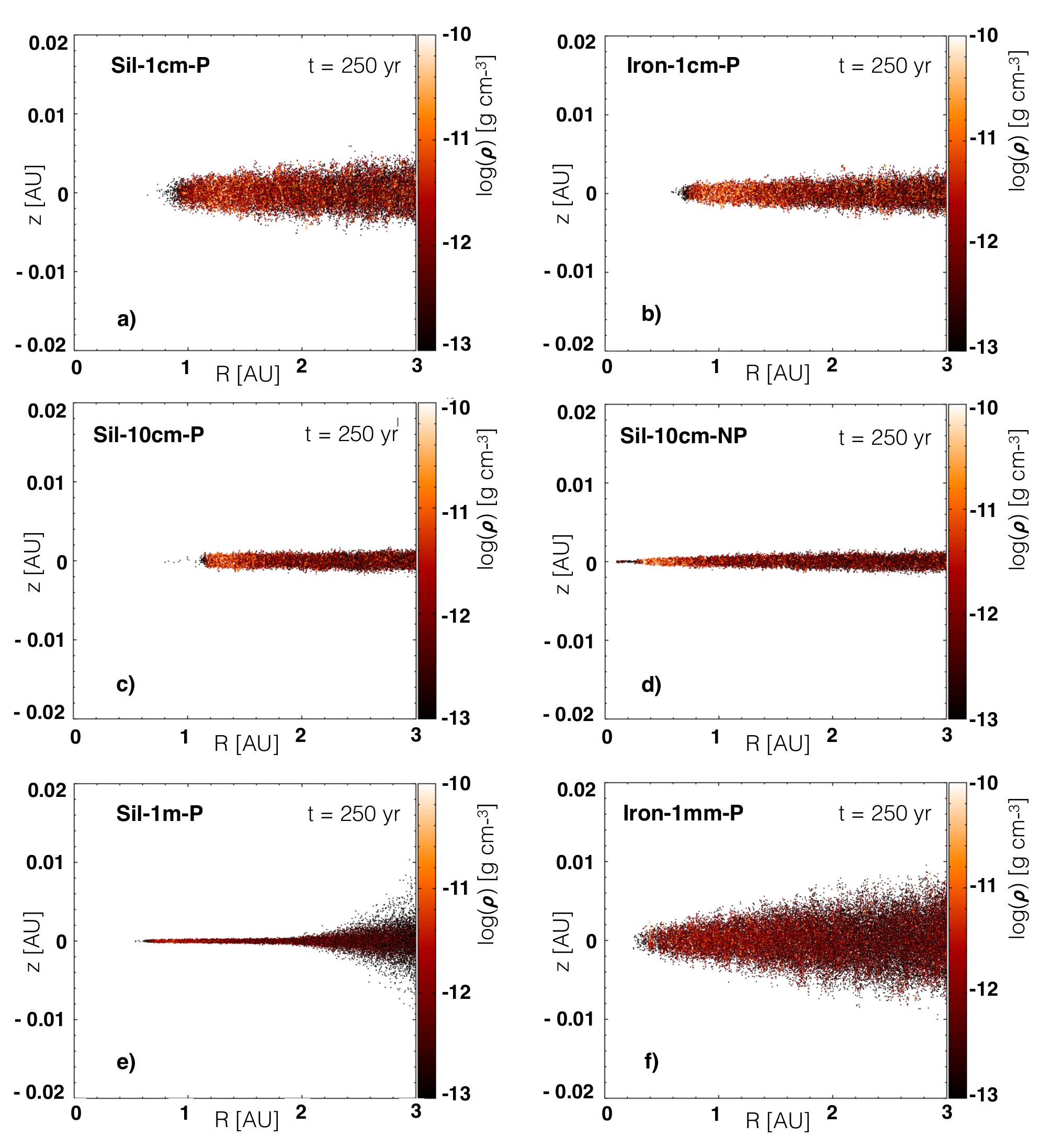}
\caption{Meridian plane cut of the dust distribution for the final state after 250 years of evolution for Sil-1cm-P (a), Iron-1cm-P (b), Sil-10cm-P (c), Sil-10cm-NP (d), Sil-1m-P (e), Iron-1mm-P (f). The coulour bar represents the logarithm of the volumetric density of the fluid.}
\label{fig:dustdiscs}
\end{figure*}	
	
\begin{table}
\begin{center}
\caption{List of P and NP simulations.}
\label{tab:simus}
\begin{tabular}{|c|c|c|c|}
\hline
Name & $\rho_d(\rm{g \, cm^{-3}})$ & $s(\rm{cm})$ & Photophoresis \\
\hline
	Sil-1cm-P & 3.2 & $10^{0}$ & ON\\
	Sil-1cm-NP & 3.2 & $10^{0}$ & OFF\\
	Sil-10cm-P & 3.2 & $10^{1}$ & ON\\
 	Sil-10cm-NP & 3.2 & $10^{1}$ & OFF\\
	Sil-1m-P & 3.2 & $10^2$ & ON\\
 	Sil-1m-NP & 3.2 & $10^2$ & OFF\\
 	Iron-1mm-P & 7.8 & $10^{-1}$ & ON \\
 	Iron-1mm-NP & 7.8 & $10^{-1}$ & OFF \\
 	Iron-1cm-P & 7.8 & $10^{0}$  & ON \\
 	Iron-1cm-NP & 7.8 & $10^{0}$  & OFF \\
\hline
\end{tabular}
\end{center}
\end{table}

\section{Results}

	\subsection{Photophoretic outward transport}

	In Fig.~\ref{fig:dustdiscs} we show the meridian plane cut of the dust distribution for different grain sizes (1~mm, 1~cm, 10~cm, 1~m) and intrinsic densities ($\rho_d^{\rm sil}$, $\rho_d^{\rm iron}$) at the end of the simulations after 250 yr. The state of the gas after the relaxation phase, shown in Fig.~\ref{fig:gasdisc}, determines the initial position of the dust particles and it is the same for all the simulations. The gas distribution remains very close to the initial state throughout the simulation. In P simulations we observe inward drift for particles located at distances larger than approximately 2 au, whereas all the NP simulations show radial drift for all the dust particles in the disc (no matter their distance to the star). This is due to the fact that dust particles close to the star feel strong photophoretic effects which result into an efficient outward transport. This phenomenon produces a sharp inner rim located at $\sim$ 1 au, which is not seen in NP simulations. The inner rim is defined as the location of the sharp decrease in the dust surface density in the inner regions of the disc. In addition, in P simulations the dust accretion by the star is completely stopped, while in NP simulations the dust is continuously accreted. Thus, photophoresis prevents the accretion by the star for particles that otherwise would have been lost. 

\begin{figure}
\includegraphics[width=90mm]{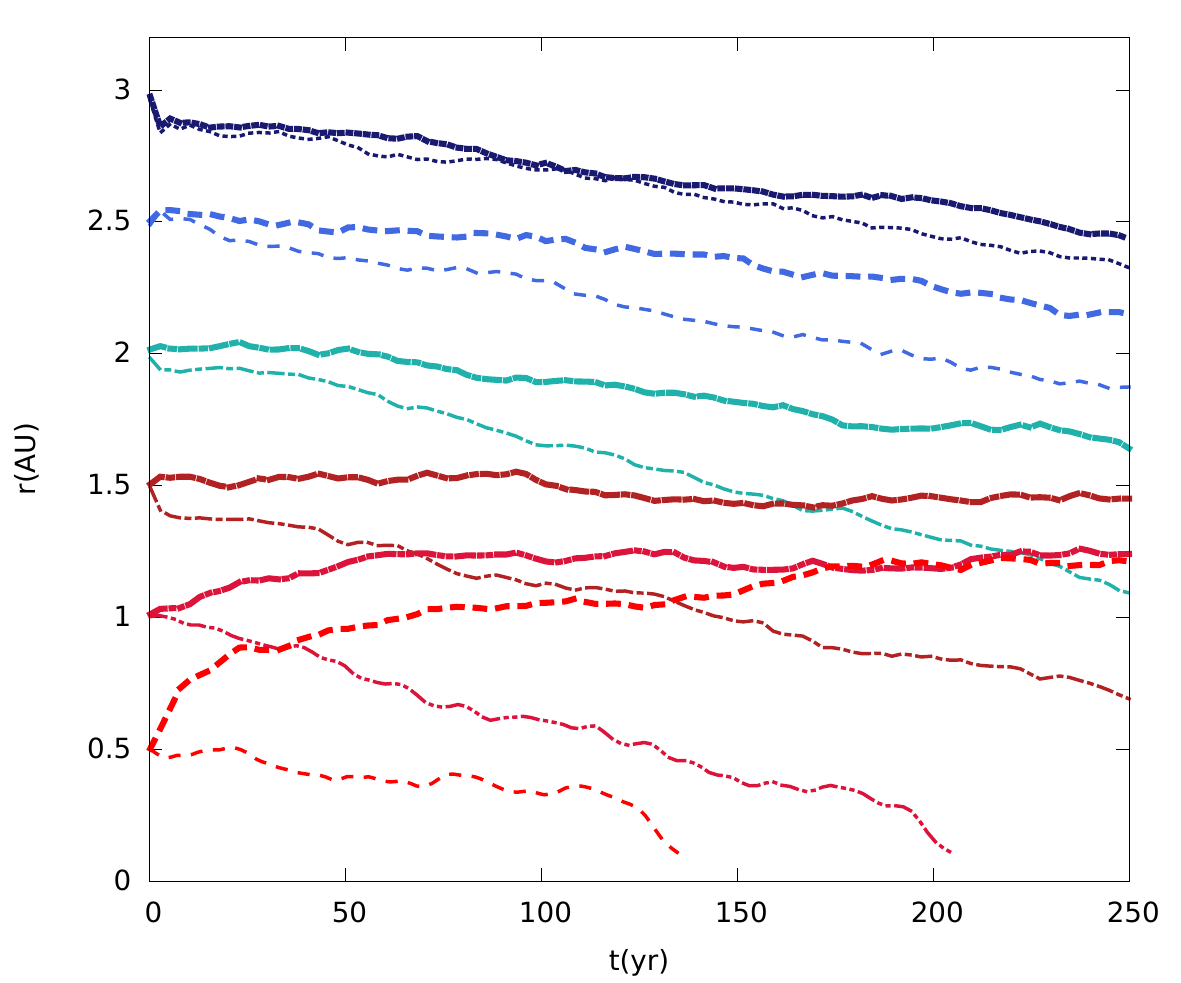}
\caption{Time evolution of the distance to the star for selected particles of different initial position Sil-10cm-P (thick lines) and Sil-10cm-NP (thin lines).}
\label{fig:traj_10cm}
\end{figure}

	In Fig.~\ref{fig:dustdiscs}d, we show the final state of the dust disc in the Sil-10cm-NP simulation to be compared with the final state of the dust disc in the simulation Sil-10cm-P in Fig.~\ref{fig:dustdiscs}c. For the former case, the inner rim is at 0.1 au and we observe a continuous dust accretion onto the star. Figure~\ref{fig:traj_10cm} shows the evolution of the radial distance for some selected particles of the simulations Sil-10cm-P and Sil-10cm-NP. Particles which are located at the same initial position have different radial motions with and without photophoresis. For $r \leq 1.5$~au, the photophoretic force on the dust particles dominates over the gas drag and the solid bodies migrate outward and accumulate at the equilibrium position; while, without photophoresis, the dust particles drift inward due to the aerodynamic drag exerted by the gas phase. These behaviours and the corresponding velocities are in agreement with previous computed radial velocities for dust particles subject to photophoresis in the solar nebula (fig.~11 in \cite{Duermann2013}).
	
		In the following discussion, Sil-10cm-P will be considered as the reference case for photophoretic transport since it shows the strongest effects (cf. Fig.~\ref{fig:dustdiscs}). We observe three types of behaviors for particles in P simulations:
\begin{enumerate}
	\item Outward motion for dust particles located between 0.1 and $\sim 1$ au from the star for Sil-10-cm-P. Without photophoresis, these particles tend to be accreted into the star in a few hundreds of orbits. The average outward radial velocities $\bar v_R$ range from 10 to 30 ${\rm m\,s^{-1}}$. 
	\item Stable orbits for particles at $r \sim 1.5$ au from the star for Sil-10cm-P. This effect is clearly due to photophoresis since it is not observed in NP simulations. In this case, the forces are balanced and $\bar v_R \sim 0$.
	\item Inward drift for particles in the outer regions of the disc outside of 1.5 au for Sil-10cm-P. Consequently, $\bar v_R < 0$. The radial motion is slower than the case without photophoresis indicating that photophoresis has a decreasing effect with increasing radial distance, as expected. 
\end{enumerate}
	
\begin{figure}
\includegraphics[width=90mm]{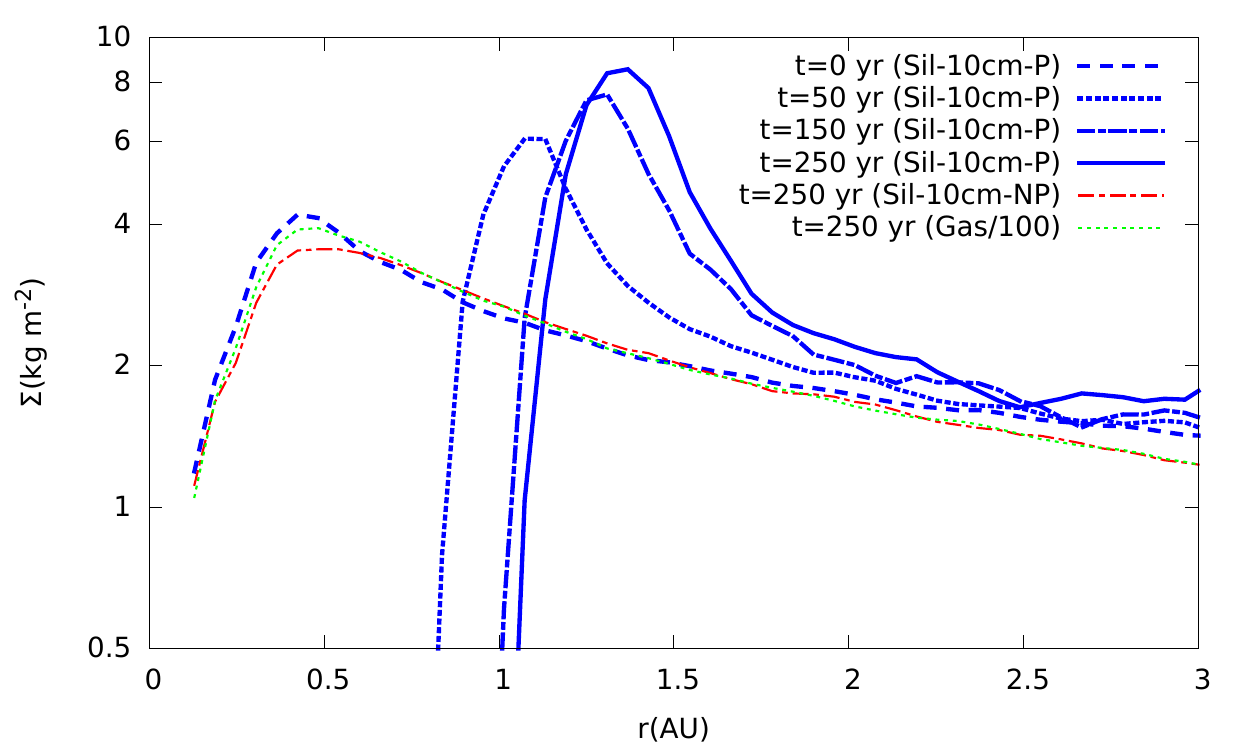}
\caption{Dust surface density for Sil-10cm-P at different evolutionary times with photophoresis (thick lines), Sil-10-cm-NP at $t=250$ yr (thin dot dashed line) and gas surface density divided by 100 at $t=250$ yr (thin dotted line).}
\label{fig:sigmavstime10cm}
\end{figure}

	In Fig.~\ref{fig:sigmavstime10cm}, we show the time evolution of the surface density for 10~cm silicate grains. With increasing time, the position of the peak moves outwards and the peak value increases. This accumulation is due to the combined effect of the outward motion of particles interior to the equilibrium position and to the inward drift of particles beyond this location. This leads to an efficient pile-up of dust at a given radial distance. The peak position evolves little with time after $t=250$ yr and is located at $\sim$ 1.3 au at $t=250$~yr. The increase of the peak value of the dust surface density is observed for all the simulations with photophoresis as shown in Fig.~\ref{fig:sigma}. The peak of the dust surface density for \mbox{Sil-10cm-NP} in Fig.~\ref{fig:sigmavstime10cm} remains very close to the peak of the gas surface density divided by 100. In fact, in the absence of photophoresis, dust accumulates at the pressure maximum of the gas distribution. The value of the peak of the dust distribution in Sil-10cm-NP is lower than the initial value because of the dust accretion.
	
 	\subsection{Accumulation regions for different grain sizes and densities}
 
 	We expect particles to accumulate at the equilibrium position at which the drag and photophoretic forces cancel each other out. In Table.~\ref{tab:rimlocation} we report the inner rim locations, $r_{\rm rim}$, and the peak position in the dust surface density, $\Sigma_{\rm d}^{\rm peak}$, for different grains species and sizes. For a given dust species, the slight discrepancy between $r_{\rm rim}$ and $\Sigma_{\rm d}^{\rm peak}$ is due to the fact that the maximum accumulation does not occurs at the rim itself.
	
\begin{figure}
\includegraphics[width=90mm]{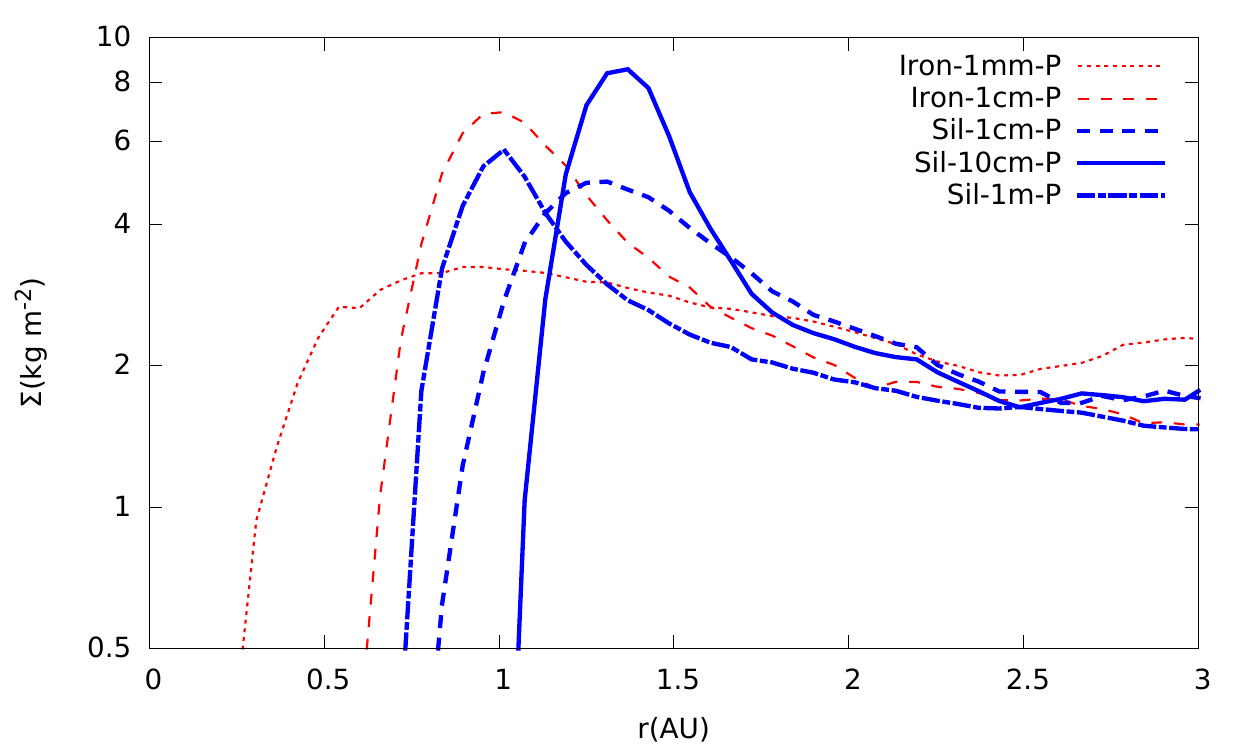}
\caption{Dust surface density at $t=250$ yr for P simulations.}
\label{fig:sigma}
\end{figure}

	The drag force in the Epstein regime is proportional to $\rho_{\rm d} s$, whereas the photophoretic force solely depends on $s$. In the radial direction, the drag force points towards the star, while the photophoretic force points in the opposite direction. Therefore, if we consider two particles of the same size but different intrinsic densities, the denser particle feels a stronger drag force compared to the lighter one. Consequently, the accumulation region for denser particles is closer to the star compared to lighter particles. This is in agreement with the inner rim locations and $\Sigma_{\rm d}^{\rm peak}$ of Sil-1cm-P and \mbox{Iron-1cm-P} reported in Table \ref{tab:rimlocation} and Fig.~\ref{fig:sigma}.
	
	For particles with the same intrinsic density but different size (1~cm, 10~cm, 1~m), analytical calculations in one-dimensional disc models show that larger particles accumulate closer to the star compared to smaller ones, as shown in fig.~1 of \cite{TakeuchiKrauss2008}. In this sense, the inner rim of Sil-1cm-P is in disagreement with this statement. However, the authors also estimate the time that it takes for a given particle to reach its equilibrium location, $\tau_{\rm mig}$, and found that it increases with decreasing size. For example, according to their calculations, for 10~cm particles $\tau_{\rm mig} \sim 100$ yr, while for 1~cm $\tau_{\rm mig} \sim 5000$ yr. In parallel, \cite{Moudensetal2011} solved the equation for the radial velocity for particles of different particle sizes in 1+1D turbulent discs and found that 10 cm particles present the largest radial velocities when compared to smaller particles. This implies that the former reach the equilibrium location on shorter time-scales (in agreement with \cite{TakeuchiKrauss2008}). After a few Myr smaller particles reach their respective equilibrium distances, which are larger than for 10~cm particles. This suggests that in our simulations mm- and cm-sized particles have not reached their equilibrium location yet. Unfortunately, the high computational cost of hydrodynamical simulations prevent us to extend this study for longer times. Nevertheless, our results correspond to an evolutionary stage of the disc for which 10~cm particles have already reached their equilibrium location and smaller grains are slowly migrating outwards. The dynamics involved in this process have important implications for the chemical separation of particles in the disc, which are discussed in Section~5.
 
\begin{table}
\begin{center}
\caption{Summary of the results at $t=250$ yr.}
\label{tab:rimlocation}
\begin{tabular}{|c|c|c|}
\hline
Simulation & $r_{\rm rim}$ (au) & $\Sigma_{\rm d}^{\rm peak}$ (au) \\
\hline
	Sil-1cm-P & 0.8 & 1.25 \\
	Sil-10cm-P & 1.1 & 1.3 \\
	Sil-1m-P & 0.6 & 1.0 \\
	Iron-1mm-P & 0.3 & 0.7 \\			
	Iron-1cm-P & 0.6 & 1.0 \\
\hline
\end{tabular}
\end{center}
\end{table}

	\subsection{Vertical settling}

	We observe that for silicate particles, the larger the grain size the more efficient the vertical settling is in the inner regions, as shown in \cite{Fouchet2005}. In the outer disc, 1~m particles do not settle because they are almost decoupled from the gas and remain on inclined orbits. In addition, comparing Sil-1cm-P in Fig.~\ref{fig:dustdiscs}a and Iron-1cm-P in Fig.~\ref{fig:dustdiscs}b, we also see that the vertical settling is more efficient for the denser particles. In fact, particles with the same product $\rho_{\rm d} s$ have the same Stokes number, i.e. dynamical behaviour. Thus, a 1~cm iron particle feels the gas drag in the same way as a silicate particle of size equal to $\rho_d^{\rm iron}/\rho_d^{\rm sil} \approx 2.44$~cm. This justifies the stronger settling observed for 1~cm iron particles compared to 1~cm silicate particles. Despite the fact the photophoretic force varies with height, we do not observe differential vertical transport due to photophoresis. In fact, if we consider particles at a fixed radial distance with different heights, they all settle towards the midplane in a few orbital periods and then vertical photophoretic effects are suppressed.

\section{Discussion}

	\subsection{Photophoretic accumulation region}

\begin{figure}
\includegraphics[width=90mm]{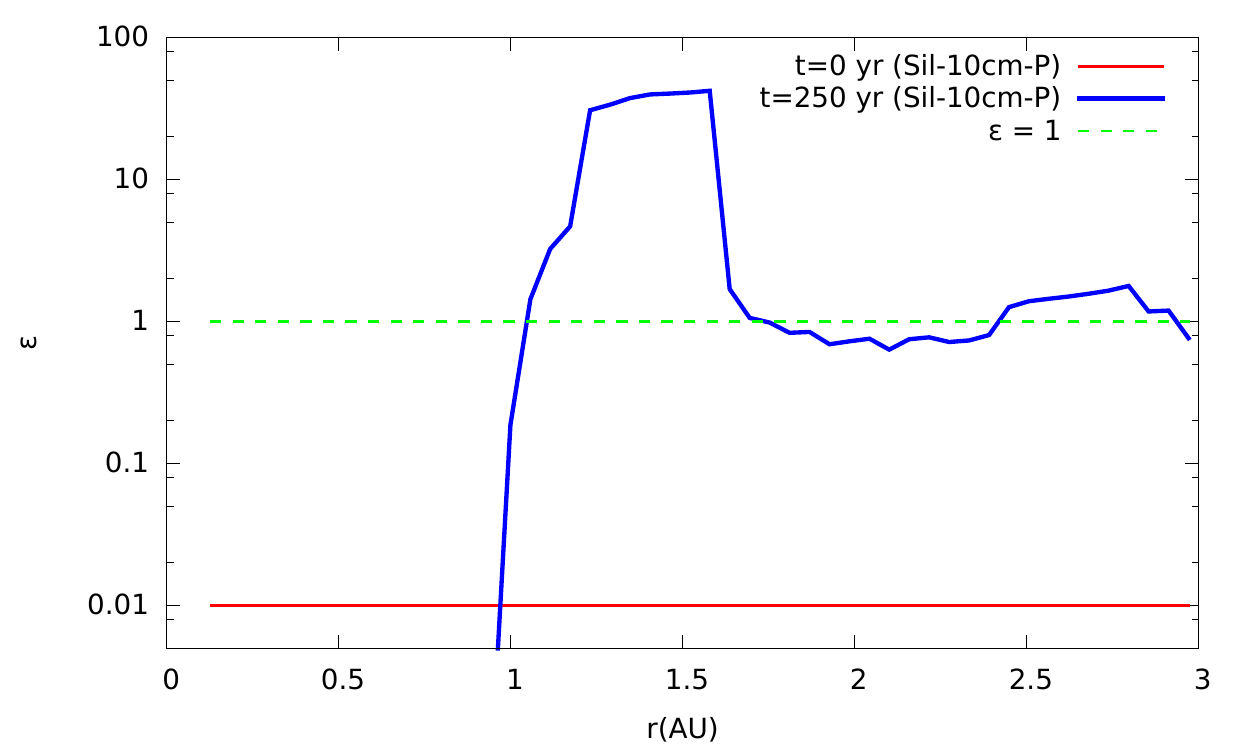}
\caption{Dust-to-gas ratio in the disc midplane for Sil-10cm-P.}
\label{fig:dusttogas}
\end{figure}

	Our results indicate that photophoresis can transport dusty grains from the inner region of the disc to the outer region with different efficiency according to the grain density and size. The dust accumulation occurs in the optically thin regions of the disc and it is due to the combined effect of stellar radiation and momentum exchange between dust and gas particles. The peak value of the surface density profiles, $\Sigma_{\rm d}^{\rm peak}$ dramatically increases in time-scales of the order of 100 yr for cm-sized particles (cf. Fig.~\ref{fig:sigmavstime10cm}) and longer time-scales for smaller particles. For example, for 10 cm silicates particles $\Sigma_{\rm d}^{\rm peak}(t = 150 \,{\rm yr}) / \Sigma_{\rm d}^{\rm peak}(t= 0 \,{\rm yr}) \approx 2$. We define the dust-to-gas ratio, $\epsilon$, as the ratio between the dust and gas volume densities in the midplane. In Fig.~\ref{fig:dusttogas} we plot $\epsilon$ as a function of the distance to the star for Sil-10cm-P. At the beginning of the simulation $\epsilon = 0.01$ by definition, and we expect $\epsilon$ to become close to 1 in the midplane with time evolution due to the vertical settling of dust \citep{Fouchet2005}. However, we observe that in the inner regions photophoretic effects results into a dramatic increase of $\epsilon$ which reaches values close to 40 between 1 and 1.7 au. In the outer parts $\epsilon$ remains close to 1 because photophoresis has little effect at large distances. The effective pile-up of dust caused by photophoresis results into favourable conditions for grain growth in the inner regions of the disc. Consequently, solid bodies may rapidly grow at this accumulation region and decouple from the gas in shorter time-scales.

	The dust accumulation caused by photophoresis is a process which differs from the particle traps discussed earlier (cf. Section 1) where a pressure maximum is required in order to collect dust at a given location in the disc. Photophoretic accumulation regions are defined as the location where drag and photophoretic forces cancel each other out. Interestingly, in our disc model, photophoresis prevents the accretion by the star for particles ranging from millimetre to metre sizes. This is particularly relevant since these particles are expected to be subject to a strong radial drift, which compromises their survival around the star. However, other disc models would show accumulation regions at different radial distances. For instance, the optically thin region in a younger and more massive disc is much less radially extended. Consequently, photophoresis will only have an observable effect on the particles very close to the gas inner edge. In this configuration, the particle survival will still hold but the size and density sorting will operate over shorter length scales. However, regardless of the model considered, our results strongly suggest that photophoresis can effectively break the radial-drift barrier in the optically thin regions of PPD.

	\subsection{Dust chemical composition}
	
\begin{figure}
\includegraphics[width=90mm]{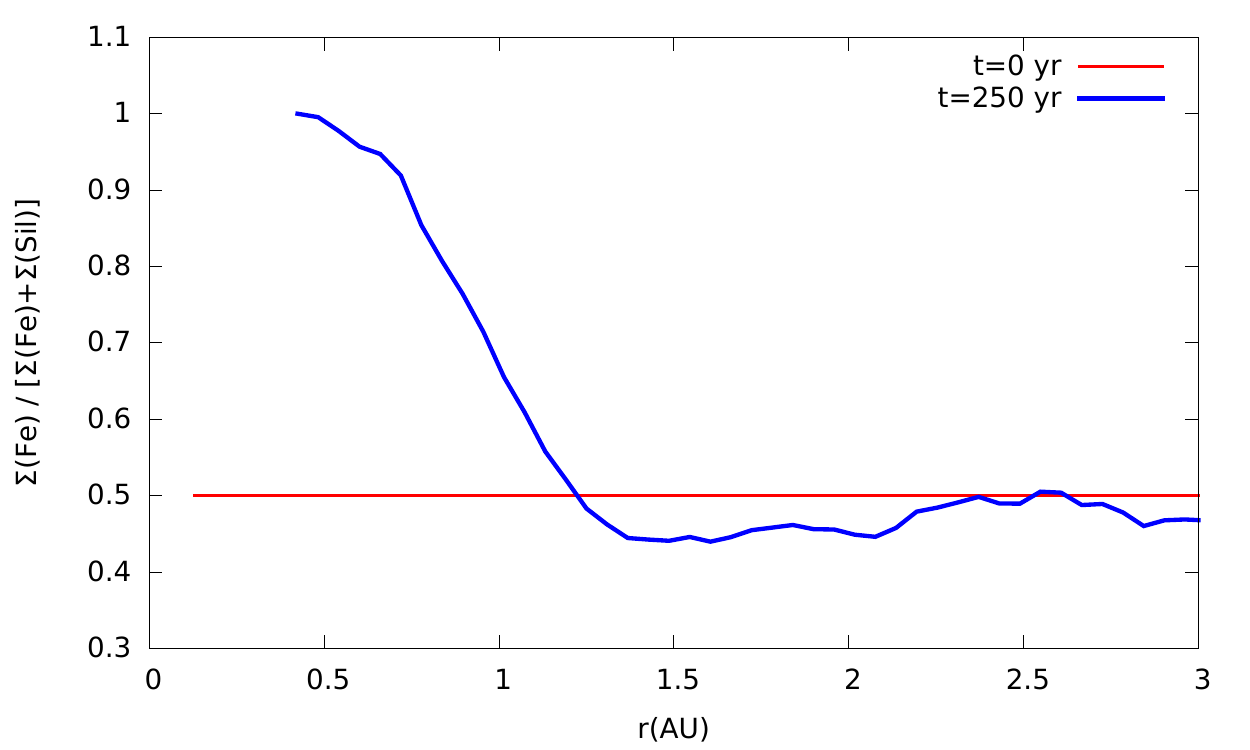}
\caption{Fe/(Fe+Sil) ratio for 1cm-sized particles with photophoresis.}
\label{fig:fesil}
\end{figure}

	In Fig.~\ref{fig:sigma} we report the dust surface density profiles at the end of the simulation for Sil-(1cm, 10cm, 1m)-P and Iron-(1mm, 1cm)-P which shows the difference between iron and silicate particles. In addition, in Fig~\ref{fig:fesil} we show the $\Sigma_{\rm d}^{\rm Fe}/(\Sigma_{\rm d}^{\rm Fe}+\Sigma_{\rm d}^{\rm Sil})$ ratio as a function of the radial distance for cm-sized particles in P simulations. We assume that, at the beginning of the simulation, the disc is made of an homogeneous mixture of iron and silicates particles, which gives $\Sigma_{\rm d}^{\rm Fe}/(\Sigma_{\rm d}^{\rm Fe}+\Sigma_{\rm d}^{\rm Sil})=0.5$ at $t=0$ yr. After 250 years of evolution, we see that the ratio becomes close to 1 in the inner regions indicating an almost iron-pure composition, while in the outer parts it is close to 0.5. From 0.4 to 1.4~au, the ratio decreases because of the increasing amount of silicate particles. This is basically due to the fact that silicate particles are more efficiently transported outwards by photophoresis compared to iron particles. Between 1.2 and 2.2~au the ratio is lower than 0.5 because this is where 1~cm silicate accumulate (cf. Fig~\ref{fig:sigma}). It has to be noted that we are only showing two species: iron as the denser extreme case and silicates as the most abundant dust species within the snow-line. We do not consider ice particles in this study since we assume to be interior to the snow-line in the region of the disc considered. In a model with more grain species, we would then expect a gradual separation between lighter and denser particles.

	The chemical sorting between iron and silicate particles shown in Figs.~\ref{fig:sigma} and \ref{fig:fesil} results in important variations in the composition in the first 2 au from the star. We demonstrate that photophoresis differentially transports grains of different intrinsic density, i.e. different chemical composition. Moreover, for a given density, we see that grain size also affects the efficiency of photophoresis (cf. Fig.~\ref{fig:sigma}). As a consequence, photophoresis may lead to size and density sorting of grains. Such a sorting is generally observed in carbonaceous chondrites where iron-rich grains are  smaller than silicate-rich chondrules \citep{Kuebler1999}. Thus, we suggest that photophoresis can be one of the possible mechanisms responsible for grain sorting such as those found in chondrites. Provided that these solids are then incorporated into large planetesimals, \cite{Wurmetal2013} showed that photophoresis can naturally explain the measured variations in the bulk chemical composition of the rocky bodies of the Solar System with pure metal grains close to the star and silicate particles farther out.
	
	\subsection{Outward transport of high-temperature forming material}

	Calcium- and aluminium-rich inclusion (CAIs) and amoeboid olivine aggregates (AOAs) are submm- to cm-sized refractory inclusions characterized by a lack of volatile elements. They are thought to be the products of high-temperature processes occurring in the inner hot regions of the solar nebula such as condensation, evaporation and melting \citep{ScottKrot2014}. However, these inclusions are observed in carbonaceous chondrites mixed with chondrules and amorphous material, presumably formed at lower temperatures \citep{MacPherson2012}. It has been suggested that an efficient chemical mixing occurred at some evolutionary stage of the PPD. In addition, the \textit{Stardust} mission recently returned grains from Comet 81P/Wild 2 with high-temperature minerals, such as forsterite and enstatite \citep{Brownlee2006}. This indicates that high-temperature forming material formed in the inner regions has to be transported towards the outer parts of the solar nebula in order to be incorporated into larger bodies.
	
	\cite{Mousisetal2007} showed that photophoresis can result into an influx of refractory material to the outer regions of the disc. Our 3D simulations show an efficient radial transport of particles located close to the star ($\sim 0.5$ au) to distances beyond 1 au, where meteorite parent bodies are thought to have formed \citep{ScottKrot2014}. In \cite{Mousisetal2007}, the particles of the inner regions are transported even farther out since the authors consider disc models with larger inner cavities (1 and 2 au). Hence, our simulations including photophoresis are in agreement with the scenario where ``hot minerals'' are injected in the outer parts of the disc and mixed with solids formed at lower temperatures.
	
\section{Conclusion}

Photophoresis is a physical process which transports illuminated particles embedded in gaseous environments away from the radiation source. In this study, we have quantified the effects of photophoresis on the inner regions of transitional discs via 3D hydrodynamical simulations. Our results show that photophoresis dramatically affects the structure of the dusty disc, while the gaseous disc remains unchanged. We have considered grains with different sizes (1~mm to 1~m) and intrinsic densities ($\rho_d^{\rm sil} = 3.2$~g~cm$^{-3}$, $\rho_d^{\rm iron} = 7.8$~g~cm$^{-3}$). In our simulations, the dust piles up, due to photophoresis opposing radial drift, at different radial distances from the star and it depends on the grain size and intrinsic density. The accumulation of dust occurs in time-scales of the order of 100 yr for \mbox{cm-sized} particles and longer time-scales for smaller particles. In addition, the average radial velocities obtained are in agreement with \cite{Duermann2013}. This leads to a size and density sorting of the dusty material in the inner regions. Hence, photophoretic effects in the disc result into the formation of axisymmetric accumulation regions around the star, which correspond to the ring-shaped features first proposed by \cite{KraussWurm2005}. In this context, photophoresis is a promising mechanism to break the radial-drift barrier and maintains dusty material in the disc inner regions.
	 
	 Our results also show that photophoresis produces an important chemical sorting of dust on the radial direction. It is worth highlighting that the vertical settling of the species of different size and density results into chemical and size variations in the vertical direction as well. Consequently, the combined effect of photophoresis and vertical settling produces a size and chemical distribution which varies with distance and height in the optically thin regions of the disc. Our findings also support the scenario where high-temperature forming material, likely formed of high-temperature minerals like enstatite and forsterite, is transported towards the outer regions of the disc.
	 
	 However, it remains unclear how the outward radial transport affects the chemical composition of the resulting solid bodies. In fact, one of the questions raised by our study is if solids can grow fast enough at the accumulation regions in order to capture and acquire the local chemical composition of the disc. If this is the case, the resulting bulk chemical composition might also vary with the height above the midplane. It is hard to predict a priori the resulting chemical composition at each location of the disc due to the complex process of grain growth. Indeed, the outcome of grain collisions strongly depends on the composition, structure and relative velocity of the colliding bodies \citep{BlumWurm2008, Wada2009, Blum2010, Zsom2010, Wada2013}. The study of grain growth in the optically thin regions of the disc taking into account photophoretic effects will be addressed in detail in future work.

\section*{Acknowledgements}
We thank the anonymous referee for useful comments. We also thank Gerhard Wurm and Markus Kuepper for enlightening discussions about photophoresis experiments.  This research was supported by the Programme National de Physique Stellaire and the Programme National de Plan\'etologie of CNRS/INSU, France. The authors are grateful to the LABEX Lyon Institute of Origins (ANR-10-LABX-0066) of the Universit\'e de Lyon for its financial support within the program ``Investissements d'Avenir" (ANR-11-IDEX-0007) of the French government operated by the National Research Agency (ANR). All the computations were performed at the Common Computing Facility (CCF) of the LABEX LIO. Figs.~\ref{fig:gasdisc} and \ref{fig:dustdiscs} were made with SPLASH \citep{Price2007}.

\bibliographystyle{mnras}
\bibliography{photobiblio}

\end{document}